\documentclass{article}
\usepackage{graphicx} 

\title{Benchmarking Linux CGroup and Schedule Idle }
\author{Anna Arpaci-Dusseau}
\date{December 2023}
\usepackage{biblatex} 
\usepackage{outlines}
\usepackage[font=small,labelfont=bf,tableposition=top]{caption}
\DeclareCaptionLabelFormat{andtable}{#1~#2  \&  \tablename~\thetable}

\usepackage{float}

\begin{document}

\maketitle

\section{Introduction}

cgroups are a Linux kernel primitive that limits, accounts for, and isolates the resource usage (CPU, memory, disk I/O) of a collection of processes. Many provider-side serverless systems (Mitosis \cite{rdma}, gVisor \cite{gvisor}, Havest VM \cite{harvest}, FAASM \cite{faasm}) leverage linux containers based around cgroups/namespaces for their performance isolation. However, how well does this primitive map to the needs of serverless platforms? How does the linux CFS scheduler deal with cgroups and SCHED\_IDLE and what are the core sources of high tail latency? Would a new primitive design fit the needs of serverless isolation better?

A first step in answering these questions is to identify the current behavior of cgroups and a sched\_idle policy via a thorough benchmarking study. We emulate the setting of latency-critical (LC) server in the presence of an intensive background (BE) process and note the impact of each primitive on the performance of the server. We show it is non-trivial to balance the potential tail-latency impact on a LC server with the throughput of BE tasks. Futhermore, there are many variables to adjust for the developer between hard maximum bandwidth limits, weighted cpu shares, the presence of the sched\_idle policy, or additionally "niceness" priority in the scheduling policy.

There has been some work exploring changes to CFS and cgroups to improve serverless performance \cite{CFS-aware}, but little in the space of how this would interact with the sched\_idle policy or different priority tasks.

The complete code for the cgroup-benchmark suite is available at:

\texttt{https://github.com/AnnaAD/cgroup-benchmark}.

\section{Background}

\subsection{Cgroups and CFS}

\textbf{cpu.weight} controls the proportional distribution of cpu time for each cgroup. \cite{website:redhat-cgroup}
For example: 
$$\texttt{echo "150" > /sys/fs/cgroup/Example/g1/cpu.weight}$$
$$\texttt{ echo "100" > /sys/fs/cgroup/Example/g2/cpu.weight}$$
$$\texttt{echo "50" > /sys/fs/cgroup/Example/g3/cpu.weight}$$ 
This command set the weights of the children cgroups (g1, g2, g3). These weights are then summed up to a total cpu weight. The CPU resource is then distributed proportionally based on the that group's individual weight compared to the sum of all weights. As a result, when all processes run at the same time, the kernel allocates to each of them the proportionate CPU time based on their respective cgroup’s cpu.weight.

\textbf{cpu.max} sets the maximum bandwidth for all processes within a cgroup. \cite{website:redhat-cgroup} For example: $$\texttt{echo "200000 1000000" > /sys/fs/cgroup/Example/tasks/cpu.max}.$$ The first value is the allowed time quota in microseconds for which all processes collectively in a child group can run during one period. The second value specifies the length of the period. During a single period, when processes in a control group collectively exhaust the time specified by this quota, they are throttled for the remainder of the period and not allowed to run until the next period. 

\subsection{SCHED\_IDLE}

The CFS (completely fair scheduling) class hosts most of the user tasks; it implements three scheduling policies: SCHED\_NORMAL, SCHED\_BATCH, and SCHED\_IDLE. \cite{website:lwn}, \cite{website:redhat-idle} A task under any of these policies gets a chance to run only if no other tasks are enqueued in the deadline or realtime classes (though by default the scheduler reserves 5\% of the CPU for CFS tasks regardless). The scheduler tracks the virtual runtime (vruntime) for all tasks, runnable and blocked. The lower a task's vruntime, the more deserving the task is for time on the processor. CFS accordingly moves low-vruntime tasks toward the front of the scheduling queue.

The priority of a task is calculated by adding 120 to its nice value, which ranges from -20 to +19. The priority of the task is used to set the weight of the task, which in turn affects the vruntime of the task; the lower the nice value, the higher the priority. The task's weight will thus be higher, and its vruntime will increase more slowly as the task runs.

In Linux kernel version 5.4, the CFS scheduler behavior was modified to favor running new SCHED\_NORMAL tasks on a warm SCHED\_IDLE cpu, fixing previous bug in SCHED\_IDLE which would have an idle task running on a single CPU for prolonged periods of time \cite{website:lwn}

\section{cgroup-benchmark suite}

The aim of cgroup-benchmark suite is to simulate a LC server with a memory intensive BE background task. We select Matrix Multiplication as our memory intensive task, and implement our multi-threaded TCP server in C++ with the \texttt{<threads>} library.

\begin{itemize}
    \item BE Matrix Multiplication: \texttt{matmul.py <nrows> <ncols>} - multiplies two $\texttt{nrows} \times \texttt{ncols}$ dimension matrices together using numpy. Numpy implements a multithreaded matrix multiplication function.
    \item LC Server \texttt{multi\_tcp\_server/server <mm?> <ncols> <nrows>} - a multithreaded TCP server which creates a new thread per client. If \texttt{mm = TRUE}, then on recieving a client request the server performs a single matrix multiplication of two $\texttt{nrows} \times \texttt{ncols}$ dimension matrices before responding. If \texttt{mm = FALSE}, the server responds right away.
    
    \item LC Server Client - a simple process which repetedly pings the server and records the obeserved RTT of each ping.
\end{itemize}

\subsection{Attempted Configurations}
\begin{outline}
    \1 Scenario 1: Light Server Load -- Server has no workload, responds immediately to any client requests.
           \2 Baseline: Start LC server and 16 clients. BE process of 500x500 matrix multiplication begins after 30 seconds.
        \2 cgroup-weight: Start LC server and 16 clients in cgroup \texttt{server}.  BE process of 500x500 matrix multiplication in cgroup  \texttt{background} begins after 30 seconds. cgroups are weighted \texttt{server}:\texttt{background}, \texttt{2048}:\texttt{1024}. 
        \2 cgroup-max: Start LC server and 16 clients in cgroup \texttt{server}.  BE process of 500x500 matrix multiplication in cgroup  \texttt{background} begins after 30 seconds. cgroup \texttt{server} is configured to have no maximum bandwidth, while \texttt{background} has a maximum of 50\% CPU bandwidth.
        \2 cgroup-sched\_idle-weight: Start LC server and 16 clients in cgroup \texttt{server}. BE process of 500x500 matrix multiplication in cgroup  \texttt{background} with \texttt{SCHED\_IDLE} policy begins after 30 seconds.
        \2 cgroup-sched\_idle-max: Start LC server and 16 clients in cgroup \texttt{server}. BE process of 500x500 matrix multiplication in cgroup  \texttt{background} with \texttt{SCHED\_IDLE} policy begins after 30 seconds. cgroups are weighted \texttt{server}:\texttt{background}, \texttt{2048}:\texttt{1024}.
        \2 sched\_idle: Start LC server and 16 clients. BE process of 500x500 matrix multiplication begins after 30 seconds with \texttt{SCHED\_IDLE} policy. cgroup \texttt{server} is configured to have no maximum bandwidth, while \texttt{background} has a maximum of 50\% CPU bandwidth.
    \1 Scenario 2: Heavy Server Load -- Server has heavy workload, performs a 500x500 malloced matrix multiplication upon receiving a client request.
        \2 Baseline
        \2 cgroup-weight
        \2 cgroup-max
        \2 cgroup-sched\_idle-weight
        \2 cgroup-sched\_idle-max
        \2 sched\_idle
    \1 Scenario 3: Reasonable Server Load  -- Server has reasonable workload, performs a 100x100 matrix multiplication upon receiving a client request.
        \2 Baseline
        \2 cgroup-weight
        \2 cgroup-max
        \2 cgroup-sched\_idle-weight
        \2 cgroup-sched\_idle-max
        \2 sched\_idle

\end{outline}

\section{Evaluation Results}

The experiments were run on Cloudlab 2-node LAN experiment configuration. Each node has 16 cores. Linux Kernel Version: 5.15.0-86-generic (which uses the CFS scheduler).

The below tests use cpu.weight as the means to distribute cpu time between the server and background tasks. However, tests were run with cpu.max, though cpu.weight seems to show the trends better. There are 16 clients, that the number of client is proportional to the number of cores and there are not many more cores than clients, or many fewer (which would trivially result in completely unused cores for BE processes).

\subsection{Reasonable Server Load}
The most clear, promising results, came from the reasonable server load.

\begin{figure}[H]
    \centering
    \begin{minipage}[c]{0.47\linewidth}
    \begin{tabular}{|l|l|l|}
    \hline
     & \textbf{latency (ns)} & \\ \hline 
    \textbf{P50:}    & 73563070  &   -    \\ \hline
    \textbf{P99:}    & 82408201   &  -  \\ \hline
    \textbf{P99.99:} & 88914093   & -\\ \hline
    \multicolumn{3}{c}{\textit{With BE Task}} \\ \hline
    \textbf{P50:}    & 59819570 & -18\%         \\ \hline
    \textbf{P99:}    & 107413252 & +30\% \\ \hline
    \textbf{P99.99:} & 135954554 & +50\%\\ \hline
    \end{tabular}
    \end{minipage}
    \begin{minipage}[c]{0.5\linewidth}
    \includegraphics[width=1.1\textwidth]{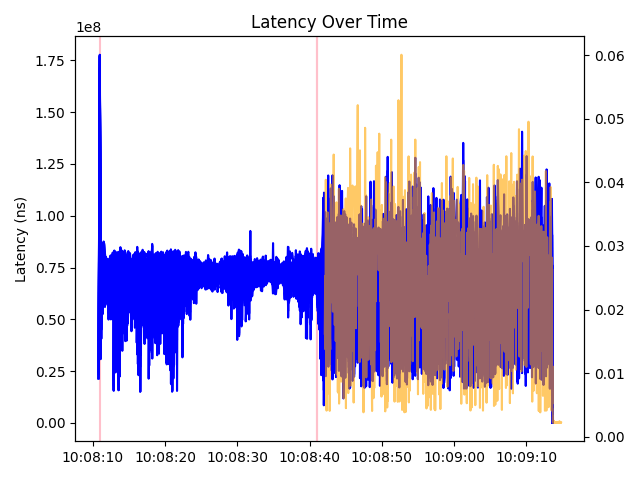}
    \end{minipage}
    \caption{Reasonable Server Load - Baseline. The blue lines indicate the latency of the 16 clients, the orange line indicates the latency of matrix multiplication for the BE task.}
    \label{fig:bug}
    \end{figure}

\begin{figure}[H]
    \centering
    \begin{minipage}[c]{0.47\linewidth}
    \begin{tabular}{|l|l|l| }
    \hline
     & \textbf{latency (ns)} & \\ \hline 
    \textbf{P50:}    & 74577791  &-   \\ \hline
    \textbf{P99:}    & 83403921  &   -  \\ \hline
    \textbf{P99.99:} & 91537968  & -\\ \hline
    \multicolumn{3}{c}{\textit{With BE Task}} \\ \hline
    \textbf{P50:}    & 71003110  & -4\%     \\ \hline
    \textbf{P99:}    & 105574298 &  +15\%     \\ \hline
    \textbf{P99.99:} & 119188803 & +30\%\\ \hline
    \end{tabular}
    \end{minipage}
    \begin{minipage}[c]{0.5\linewidth}
    \includegraphics[width=1.1\textwidth]{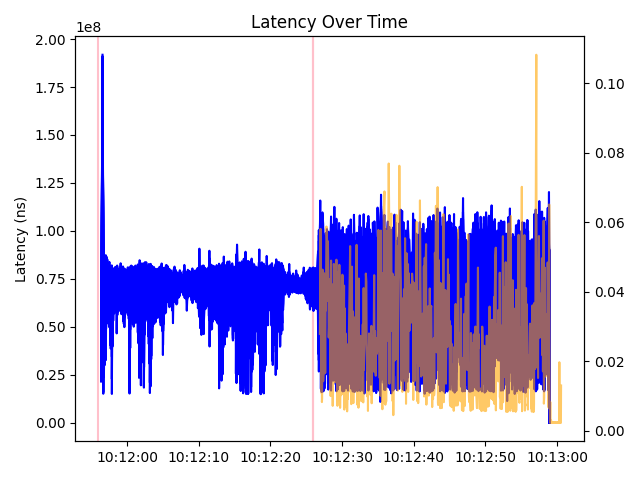}
    \end{minipage}
    \caption{Reasonable Server Load - cgroup (cpu.weight)  The blue lines indicate the latency of the 16 clients, the orange line indicates the latency of matrix multiplication for the BE task.}
    \label{fig:bug}
    \end{figure}

We see a small (12\%) decrease in 99.9\% tail latency with cgroups in this setting (with cpu.weight). Overall, with cgroups turned on, we see more stable results and a smaller effect on tail latencies when starting the BE task, with an average decrease of 20\% in the difference between tail latencies with cgroups.

\begin{figure}[H]
    \centering
    \begin{minipage}[c]{0.47\linewidth}
    \begin{tabular}{|l|l|l|}
    \hline
     & \textbf{latency (ns)} & \\ \hline 
    \textbf{P50:}    & 73738949 &   -     \\ \hline
    \textbf{P99:}    & 83778371 &   -  \\ \hline
    \textbf{P99.99:} & 92482737 &   -   \\ \hline
    \multicolumn{3}{c}{\textit{With BE Task}} \\ \hline
    \textbf{P50:}    & 66796023 &  -9\%     \\ \hline
    \textbf{P99:}    & 93924664 &  12\%     \\ \hline
    \textbf{P99.99:} & 152616914 &    +65\%    \\ \hline
    \end{tabular}
    \end{minipage}
    \begin{minipage}[c]{0.5\linewidth}
    \includegraphics[width=1.1\textwidth]{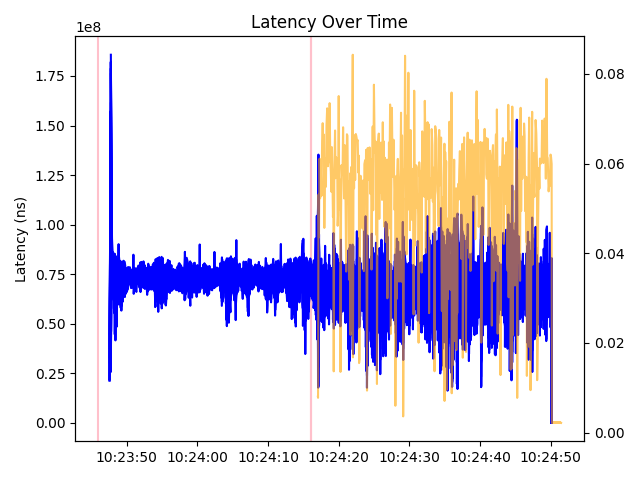}
    \end{minipage}
    \caption{Reasonable Server Load - cgroup (cpu.weight) and idle.  The blue lines indicate the latency of the 16 clients, the orange line indicates the latency of matrix multiplication for the BE task.}
    \label{fig:bug}
    \end{figure}
    
With Sched\_idle and cgroups we see slight improvement to the p99 tail latency, and it seems to change the characterisitic of the reported latencies to occur over smaller range overall. However, the p99 only improves by $\approx $5\% from without sched\_idle.

\begin{figure}[H]
    \centering
    \begin{minipage}[c]{0.47\linewidth}
    \begin{tabular}{|l|l|l|}
    \hline
    & \textbf{latency (ns)} & \\ \hline 
    \textbf{P50:}    & 73862825  & -     \\ \hline
    \textbf{P99:}    & 82168891  &  -   \\ \hline
    \textbf{P99.99:} & 103361643 & -\\ \hline
    \multicolumn{3}{c}{\textit{With BE Task}} \\ \hline
    \textbf{P50:}    & 71185558  & -3\%     \\ \hline
    \textbf{P99:}    & 82244143  & +0.1\%    \\ \hline
    \textbf{P99.99:} & 107807689 & +0.4\%\\ \hline
    \end{tabular}
    \end{minipage}
    \begin{minipage}[c]{0.5\linewidth}
    \includegraphics[width=1.1\textwidth]{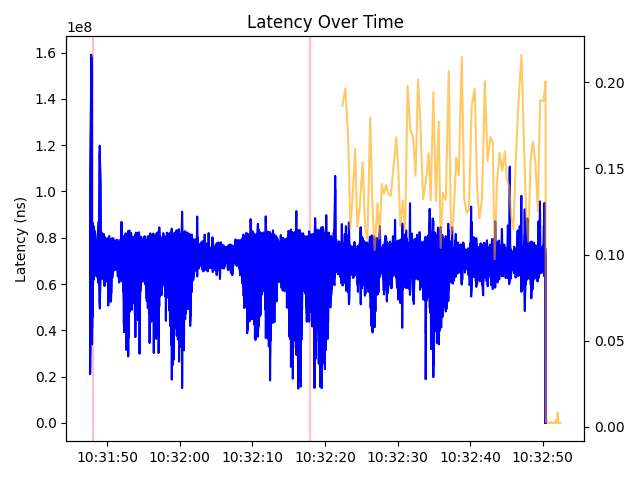}
    \end{minipage}
    \caption{Reasonable Server Load - Idle.  The blue lines indicate the latency of the 16 clients, the orange line indicates the latency of matrix multiplication for the BE task.}
    \label{fig:bug}
    \end{figure}

Using only a SCHED\_IDLE policy proved to maintain the tail latency of the server the most consistent, at the cost of decreasing the throughput of BE matrix multiplication by 7.5x.

\subsection{Light Server Load}
One possible limitation of a heavier server load is that spikes in tail latencies are less noticable overall due to a higher baseline RTT. If the work of the server before responding is much higher than any kernel overhead due to cgroups/sched\_idle we might not identify a millisecond-level increase.

One limitation of a very light server load is there seems to be performance benefits from starting a BE task, as it potentially warms and stabilizes the CPU frequency. For all these experiements, the CPU Frequency of all cores was set to the maximum (2000 mHz), however, even with this mitigation, we see that starting real work on the CPU improves performance of the LC server.

  \begin{figure}[H]
    \centering
    \begin{minipage}[c]{0.4\linewidth}
        \begin{tabular}{|l|l|}
        \hline
        & \textbf{latency (ns)}  \\ \hline 
        \textbf{P50:}    & 218037          \\ \hline
        \textbf{P99:}    & 589715  \\ \hline
        \textbf{P99.99:} & 978404  \\ \hline
        \multicolumn{2}{c}{\textit{With BE Task}} \\ \hline
        \textbf{P50:}    & 117088           \\ \hline
        \textbf{P99:}    & 250568          \\ \hline
        \textbf{P99.99:} & 1228421 \\ \hline
        \end{tabular}
    \end{minipage}
    \begin{minipage}[c]{0.48\linewidth}
    \includegraphics[width=1.1\textwidth]{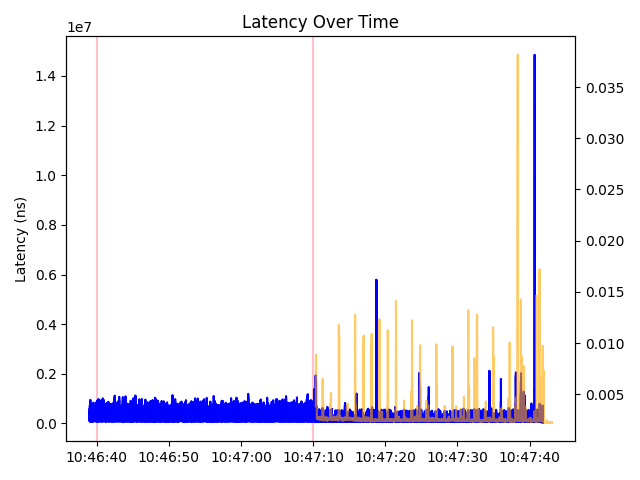}
    \end{minipage}
    \caption{Light Server Load Baseline.  The blue lines indicate the latency of the 16 clients, the orange line indicates the latency of matrix multiplication for the BE task.}
    \label{fig:bug}
    \end{figure}

\begin{figure}[H]
    \centering
    \begin{minipage}[c]{0.4\linewidth}
        \begin{tabular}{|l|l|}
        \hline
         & \textbf{latency (ns)}  \\ \hline 
        \textbf{P50:}    & 216684           \\ \hline
        \textbf{P99:}    & 592284  \\ \hline
        \textbf{P99.99:} & 1021502\\ \hline
        \multicolumn{2}{c}{\textit{With BE Task}} \\ \hline
        \textbf{P50:}    & 205964           \\ \hline
        \textbf{P99:}    & 580310  \\ \hline
        \textbf{P99.99:} & 2785662 \\ \hline
        \end{tabular}
    \end{minipage}
    \begin{minipage}[c]{0.48\linewidth}
    \includegraphics[width=1.1\textwidth]{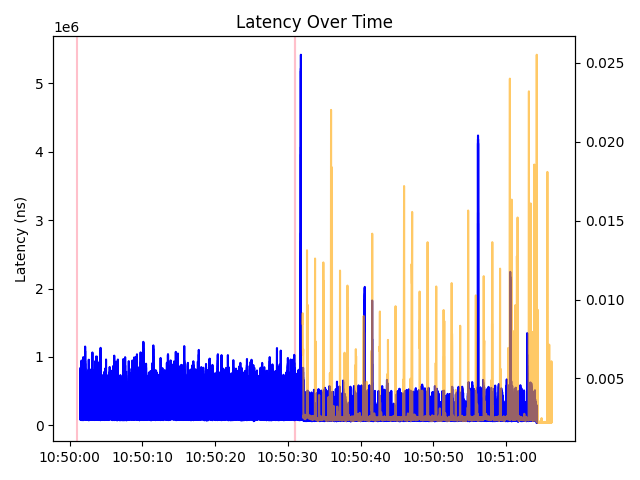}
    \end{minipage}
    \caption{Light Server Load cgroup (cpu.weight).  The blue lines indicate the latency of the 16 clients, the orange line indicates the latency of matrix multiplication for the BE task.}
    \label{fig:bug}
    \end{figure}

\begin{figure}[H]
    \centering
    \begin{minipage}[c]{0.4\linewidth}
    \begin{tabular}{|l|l|}
    \hline
     & \textbf{latency (ns)}  \\ \hline 
    \textbf{P50:}    & 217550           \\ \hline
    \textbf{P99:}    & 582401  \\ \hline
    \textbf{P99.99:} & 1025065\\ \hline
    \multicolumn{2}{c}{\textit{With BE Task}} \\ \hline
    \textbf{P50:}    & 118966          \\ \hline
    \textbf{P99:}    & 275199 \\ \hline
    \textbf{P99.99:} & 650053  \\ \hline
    \end{tabular}
    \end{minipage}
    \begin{minipage}[c]{0.48\linewidth}
    \includegraphics[width=1.1\textwidth]{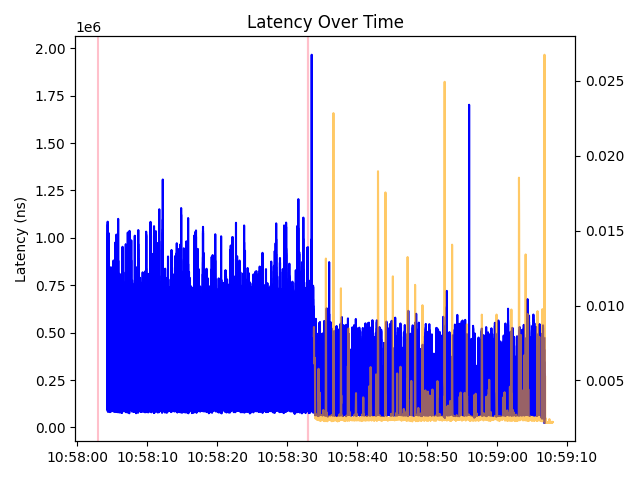}
    \end{minipage}
    \caption{Light Server Load cgroups (cpu.weight) and sched\_idle.  The blue lines indicate the latency of the 16 clients, the orange line indicates the latency of matrix multiplication for the BE task.}
    \label{fig:bug}
    \end{figure}

With sched\_idle we see a complete elimination of these most extreme tail latency spikes that occur only when the BE task is present. 

\subsection{Heavy Server Load}

The extremely heavy serverload eliminated the problem of speed-ups with BE tasks being stated, however accute spikes in tail latency were hard to detect due to a larger overall granularity in RTTs.

\begin{figure}[H]
    \centering
    \begin{minipage}[c]{0.4\linewidth}
    \begin{tabular}{|l|l|}
    \hline
     & \textbf{latency (ns)}  \\ \hline 
    \textbf{P50:}    & 2397824243           \\ \hline
    \textbf{P99:}    & 4006877900  \\ \hline
    \textbf{P99.99:} &  4021320829\\ \hline
    \multicolumn{2}{c}{\textit{With BE Task}} \\ \hline
    \textbf{P50:}    & 4503614119         \\ \hline
    \textbf{P99:}    & 5241836059 \\ \hline
    \textbf{P99.99:} & 5437727897  \\ \hline
    \end{tabular}
    \end{minipage}
    \begin{minipage}[c]{0.48\linewidth}
    \includegraphics[width=1.1\textwidth]{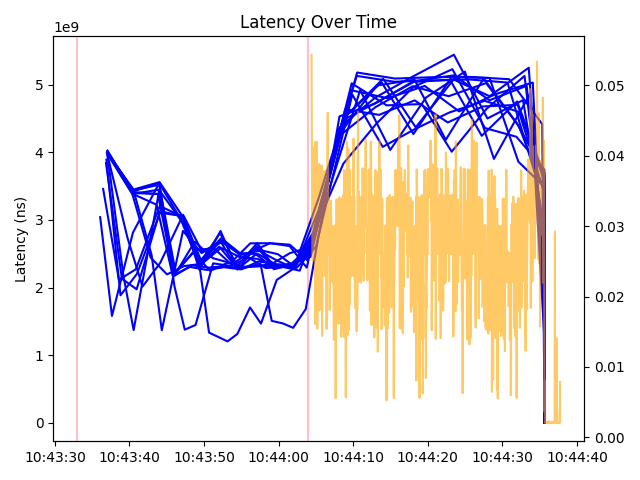}
    \end{minipage}
    \caption{Heavy Server Load Baseline.  The blue lines indicate the latency of the 16 clients, the orange line indicates the latency of matrix multiplication for the BE task.}
    \label{fig:bug}
    \end{figure}

\begin{figure}[H]
    \centering
    \begin{minipage}[c]{0.4\linewidth}
    \begin{tabular}{|l|l|}
    \hline
     & \textbf{latency (ns)}  \\ \hline 
    \textbf{P50:}    & 2397824243           \\ \hline
    \textbf{P99:}    & 4426738779  \\ \hline
    \textbf{P99.99:} &  4431107022\\ \hline
    \multicolumn{2}{c}{\textit{With BE Task}} \\ \hline
    \textbf{P50:}    &3788543978       \\ \hline
    \textbf{P99:}    & 4116269462 \\ \hline
    \textbf{P99.99:} &  4129458508  \\ \hline
    \end{tabular}
    \end{minipage}
    \begin{minipage}[c]{0.48\linewidth}
    \includegraphics[width=1.1\textwidth]{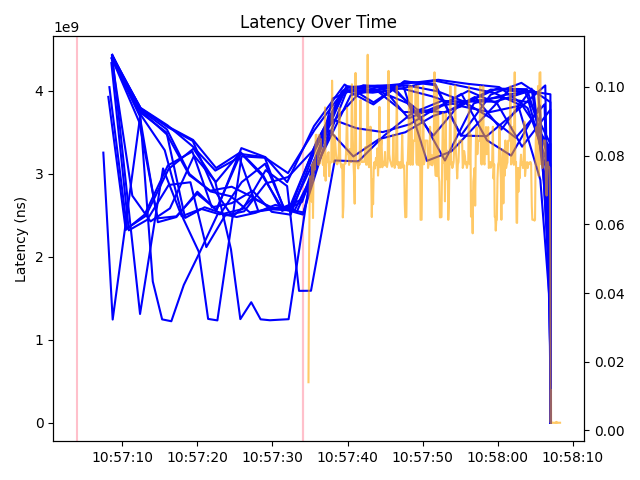}
    \end{minipage}
    \caption{Heavy Server Load with cgroups (cpu.weight).  The blue lines indicate the latency of the 16 clients, the orange line indicates the latency of matrix multiplication for the BE task.}
    \label{fig:bug}
    \end{figure}

\begin{figure}[H]
    \centering
    \begin{minipage}[c]{0.4\linewidth}
    \begin{tabular}{|l|l|}
    \hline
     & \textbf{latency (ns)}  \\ \hline 
    \textbf{P50:}    & 2617646084           \\ \hline
    \textbf{P99:}    & 4431118506  \\ \hline
    \textbf{P99.99:} &  4433799972\\ \hline
    \multicolumn{2}{c}{\textit{With BE Task}} \\ \hline
    \textbf{P50:}    & 3458744314       \\ \hline
    \textbf{P99:}    & 4200929889 \\ \hline
    \textbf{P99.99:} &  4250282000 \\ \hline
    \end{tabular}
    \end{minipage}
    \begin{minipage}[c]{0.48\linewidth}
    \includegraphics[width=1.1\textwidth]{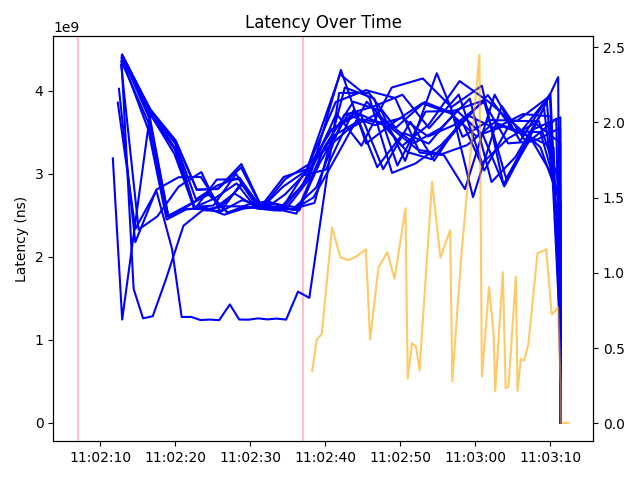}
    \end{minipage}
    \caption{Light Server Load with cgroups and sched\_idle.  The blue lines indicate the latency of the 16 clients, the orange line indicates the latency of matrix multiplication for the BE task.}
    \label{fig:bug}
    \end{figure}

Even when the tail latencies are at the coarser-grained scale, we can still see a slight improvement with sched\_idle policy. There was a 57\% in p50 latency with/without BE task with no idle policy, whereas there was only a 32\% increase with/without BE task with idle.

\subsection{Conclusion}

Early experimentation with cgroup benchmark suites indicates that sched\_idle policy is critical in reducing one-off tail latency spikes. One theory of why sched\_idle reduces these spikes is because it leads to greater affinity of server processes to a set of cores, and reduced task migration. This would lead to fewer instances where a packet for a task arrives on a different core which can cost CPU time \cite{fast-network}. However, further work is needed to complete the explanation of these tail latency spikes.

\subsection{Future Work}

Closer tracing of what is causing the tail latency spikes is necessary to complete the story of cgroups and sched\_idle in the serverless settings. Using kernel tracing tools, like trace\_cmd visualized with KernelShark\cite{kernelshark} for scheduler behavior or a flame graph of the kernel call stacks \cite{flamegraph} could be promising. 

Furthermore, it may be useful to pin certain tasks to certain CPUs so the effect of scheduling across CPU cores vs the usage within one CPU core can be identified.

Additionally, dynamically varying the number of clients, as opposed to the static-state testing, will be critical in understanding all the effects of these kernel primitives.

\printbibliography

\end{document}